\documentclass[conference]{IEEEtran} 
\IEEEoverridecommandlockouts
\usepackage{caption}

\usepackage[hyphens]{url}
\usepackage{caption}

\usepackage{multicol, blindtext}
\usepackage{float}
\usepackage{cite}
\usepackage{amsmath,amssymb,amsfonts}

\usepackage[linesnumbered,lined,boxed,commentsnumbered,ruled,longend]{algorithm2e}
\usepackage{graphicx}
\usepackage{textcomp}
\usepackage{xcolor}

\def\BibTeX{{\rm B\kern-.05em{\sc i\kern-.025em b}\kern-.08em
    T\kern-.1667em\lower.7ex\hbox{E}\kern-.125emX}}
\usepackage[utf8]{inputenc}
\usepackage[english]{babel}

\usepackage[nomargin, inline, index,  draft]{fixme}
\fxusetheme{color}

\title{A Stable Matching Assignment for Cancer Treatment Centers using Survival Analysis}

\author{Navid Seidi\\
	\normalsize Dept. of Computer Science, Missouri University of Science and Technology\\
}

\begin{document}

\maketitle
\thispagestyle{plain}
\pagestyle{plain}

\begin{abstract}
The treatment of cancer is one of the most discussed issues in the realm of contemporary public health research. One of the primary concerns of both the general public and the government is the development of the most effective cancer treatment at the most affordable price. This is due to the fact that the number of persons diagnosed with cancer increases on an annual basis. Within the scope of this project, we propose the development of a system for the recommendation of treatment centers. This system would initially select patients who posed a higher risk value, and then it would recommend the most appropriate cancer treatment center for those patients based on their income and the location where they lived using a stable matching algorithm.
\end{abstract}




\section{INTRODUCTION}
The treatment of cancer is one of the most discussed issues in the domain of modern public health research. It is projected that there will be 1,918,030 new cases of cancer and 609,360 deaths from cancer in the United States in 2022, according to~\cite{XUS}. Approximately 350 people will lose their lives to lung cancer every single day, making it the leading cause of cancer-related deaths. According to~\cite{ZUS}, the number of newly diagnosed cases in 2021 was 1,898,160, and the number of fatalities was 608,570. In addition, they demonstrated in study~\cite{YChina} that there will be approximately 4,820,000 newly diagnosed cases of cancer and approximately 3,210,000 deaths due to cancer in China by the end of 2022. In China, the majority of people who pass away from cancer do so as a result of lung cancer (\cite{YChina}). As a result of the findings presented in \cite{WChina}, which projected that there would be 4,568,754 newly diagnosed cases of cancer in China in 2020 and 3,002,899 deaths from cancer, the trend is also increasing in China. Based on these findings, it is clear that one of the primary concerns of both the people and the government is to offer the most effective treatment method at the most affordable price.\\
The National Cancer Institute Cancer Centers Program is one of the most important pillars of the United States overall cancer research effort. It was established in 1971 as a part of the National Cancer Act in order to facilitate the provision of the necessary services for the treatment of cancer. Through this program, the National Cancer Institute (NCI) recognizes centers all over the United States that meet stringent standards for conducting multidisciplinary, cutting-edge research with the goal of developing novel and more effective strategies for preventing, diagnosing, and treating cancer~\cite{NCI}. Patients in communities all over the United States can receive innovative cancer treatments at NCI-Designated Cancer Centers~\cite{NCI}.
Given that it is not possible to provide services to all of these patients in these centers, in this project, we present a novel technique for selecting the optimal center for patients who are more at risk than others, based on their place of residence and their economic level.
In this regard, we begin by determining the status of patients by first determining their relative risk score with the use of survival analysis. In the following step, we will select the patients who are eligible to receive treatment at the cancer centers based on the amount of space currently available for new patients. The third step of a stable matching algorithm involves determining the optimal center for each individual depending on the individual's current location and income. Finally, these recommendations are compiled into a single document, which is then presented to the patients and NCI for final decision-making.\\
This paper is organized as follows: in Section~\ref{RW}, we examine prior research on the Survival Analysis and Stable Matching, and in Section~\ref{GPF}, we provide a detailed description of how this problem can be formulated as a graph problem. Section~\ref{ALG} outlines the main algorithm of our research, whereas Section~\ref{Analysis} discusses the experimental phase and validations. In section~\ref{CL}, we present a conclusion, and in section~\ref{FW}, we discuss future research.

\section{Related Works}
\label{RW}

Since studies on Survival Analysis and Stable Matching are the two primary areas of this study, we will conduct a separate review of the historical data pertaining to each of them in this section.
\subsection{Survival Analysis}
The first part of the proposed project is based on calculating the risk score for each cancer patient using Survival Analysis models. This has been done in many studies. Recent research in survival analysis has mostly centered on the application of machine learning techniques to non-parametric or semi-parametric methods for predicting survival functions. Moreover, several groundbreaking works have been published that make use of neural networks.
For example, in~\cite{DeepSurv}, the authors introduced a Cox proportional hazard deep neural network and a state-of-the-art survival method for modeling interactions between a patient's covariates and treatment effectiveness in order to provide personalized treatment recommendations and to offer fully parametric models with competing risks~\cite{DeepSurv}. Deep Survival Machines (DSM) is a new, entirely parametric technique for assessing relative risks in time-to-event prediction difficulties, including censored data that is both accurate and efficient, as described in the~\cite{DSM} by the authors. Typically, it is necessary to make considerable assumptions about the constant proportional hazard of the underlying survival distribution in order to apply the Cox model, while their technique does not demand the use of significant assumptions about this variable. They demonstrate the benefits of their technique when used to predict survival risks by simultaneously learning deep nonlinear representations of the input variables from the input data through thorough testing on multiple real-world data sets with varying amounts of censoring. DeepHit~\cite{DeepHit} uses a deep neural network to directly learn the distribution of survival times rather than employing a statistical model to accomplish this. The~\cite{DeepHit} risk model, in contrast to other risk models, makes no assumptions about the underlying stochastic process and accepts the possibility that the relationship between variables and risk(s) may change over time.
This provides a concern; while the previously stated works are designed to compute the survival time and hazard function using information directly connected to each individual's health status, they do not include information about the geographical area where the patient is diagnosed with the disease or is being treated. However, many studies have revealed that the treatment environment significantly impacts the healing process. As a result, patients, their families, and healthcare professionals are becoming more aware of the importance of the physical environment in the healing process and their overall well-being~\cite{Coakley2009} the influence of pet therapy on the healing process is being studied. Patients demonstrated significant decreases in pain, respiration rate, and negative emotional State compared to baseline, as well as a significant rise in felt energy level.
Therefore, when we see that environmental factors play an essential role in the patient's recovery, the calculation accuracy of Survival Analysis could increase by adding those factors to the datasets. To achieve this goal, it is possible to add variables such as the patient's place of residence and the general health status of the patient's living place to the study. Also, other attributes such as the quality level of the treatment center and the level of knowledge, ability, and experience of the treatment staff can significantly impact more accurate calculations.
In~\cite{Xue2020}, they proposed a geographically weighted Cox regression model for sparse spatial survival data.' This research depends on having the exact geographical location of people and their distance to specific points. In spite of the presented novelty, since it is highly dependent on exact geographical coordinates, it is not applicable in the real world considering the limitations related to people's privacy and the Health Insurance Portability and Accountability Act (HIPAA).
In~\cite{Seidi2022}, a complementary improvement to survival regression models by incorporating public health statistics in the input features. We show that including geographic location-based public health information results in a statistically significant improvement in the concordance index evaluated on the Surveillance, Epidemiology, and End Results (SEER)~\cite{SEER} dataset containing nationwide cancer incidence data. Due to the improvements in calculating the risk score and also using geographic location-based data in the calculations, this method is the best one for the first part of the project.

\subsection{Stable Matching}
The second part, which is finding and recommending the best available cancer centers, can be defined as a stable matching problem. At first, this problem may seem like a simple “College Admission and the Stability of Marriage” problem that can be easily solved by the method presented in~\cite{Gale1962}, but it is not. Because, firstly, in~\cite{Gale1962} they assume “other things being equal, students should receive consideration over colleges”. This means the algorithm is student-optimal (patient-optimal), but in a treatment scenario, the situation is college-optimal (center-optimal in our case). Secondly, in~\cite{Gale1962} they assumed that “if a college is not willing to accept a student under any circumstances, then that student will not even be permitted to apply to college”. However, such an assumption is not valid in cancer treatment, because if there is any available space (bed) it will not remain empty as soon as there is an unassigned patient on the waiting list. The other significant difference in this problem is that, despite the stable matching in the Marriage or College Admission problems, patients do have not any preferences and the ordered list of considerable cancer centers, should be calculated by the algorithm.
Two other applications of this type of stable matching are “Rural Hospitals Theorem (RHT)” and “National Resident Matching Program (NRMP)”. RHT is a fundamental theorem in the theory of stable matching. It analyzes the problem of matching doctors to hospitals for residency, where each doctor is linked to a single hospital but each hospital has multiple opportunities for doctors. Because the overall number of vacancies exceeds the total number of doctors, some hospitals will undoubtedly stay empty. Rural hospitals are typically less sought after than urban hospitals, hence they frequently have several open posts. This prompted the question of whether the method used to link doctors to hospitals may be altered to assist these rural hospitals. This application has been studied in~\cite{Roth1986} and~\cite{Klijn2014}. However, the main difference between our idea and this problem is that in cancer treatment, the available admission capacity (number) is always less than the number of applicants.
NRMP as explained in~\cite{Roth2003}, commonly known as The Match, is a private non-profit non-governmental organization established in the United States that was founded in 1952 to put U.S. medical school candidates in residency training programs in teaching hospitals in the United States. Its aim has now grown to encompass the placement of international medical school patients and graduates from both the United States and other countries into residency and fellowship training programs. As described in~\cite{Mullin1951},~\cite{Almost} this problem has also been solved the same as the proposed algorithm for “College Admission and the Stability of Marriage” by~\cite{Gale1962} The latest extensions to this solution are only focused on: 

\begin{itemize}
    \item Finding a way to reduce the number of not-filled positions in colleges (medical schools) after finishing the assignments and during the decision-making time by patients to accept or reject the offers.
    \item Find a way to suggest the best algorithm for stable matching when applicants are couples.
\end{itemize}
It is obvious that none of the above are concerns in our problem, because firstly, no cancer center has empty available beds due to the lack of the number of beds. Secondly, couples with the same disease do have not a big number of cancer patients.

There are some other similar works that have been defined as a matching problem. In~\cite{Wang2018} they define stability for ride-share matches and present several mathematical programming methods to create stable or nearly stable matches, noting that ride-share matching optimization is done over time with incomplete information. In~\cite{Kakimura2021}, the Patient algorithm, which waits to thicken the market, is exponentially superior to the Greedy algorithm, which matches agents greedily. This means that waiting has substantial benefits in maximizing a matching over a bipartite network. The Patient algorithm is best when the planner identifies agents about to quit the market, they say. In~\cite{Roth2004} they investigate how larger-scale exchanges of this kind can be arranged in a way that is both efficient and incentive compatible, while still adhering to the constraints that are already in place.

In the next section, the properties of the related graph for this problem is introduced. The proposed algorithm will be run on this graph. It has two phases:
\begin{itemize}
    \item As the first step, patients with higher risk scores than the threshold will be selected.
    \item An algorithm will be run on the available bi-partite graph to find the best possible stable matching based on the available information and the thresholds.
\end{itemize}

\section{Graph-theoretic Problem Formulation}
\label{GPF}
\begin{figure}[H]
    \centering
    \includegraphics[width=0.5\textwidth]{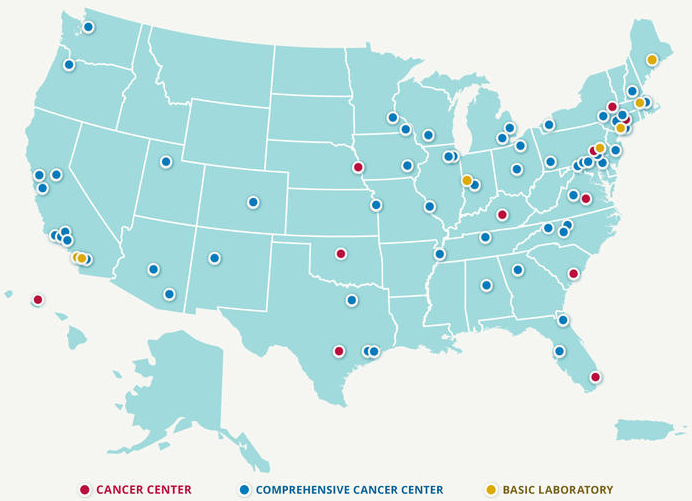}
    \caption{Types and distribution of NCI-Designated Cancer Centers across the United States~\cite{NCI}.}
    \label{fig:NCI}
\end{figure}

According to~\cite{NCI}, There are 71 NCI-Designated Cancer Centers, located in 36 states and the District of Columbia (Fig.~\ref{fig:NCI}), that are funded by NCI to deliver cutting-edge cancer treatments to patients. Of these 71 institutions:
\begin{itemize}
    \item 11 are Cancer Centers, recognized for their scientific leadership, resources, and the depth and breadth of their research in basic, clinical, and/or prevention, cancer control, and population science.
    \item 53 are Comprehensive Cancer Centers, also recognized for their leadership and resources, in addition to demonstrating an added depth and breadth of research, as well as substantial transdisciplinary research that bridges these scientific areas.
    \item 7 are Basic Laboratory Cancer Centers that are primarily focused on laboratory research and often conduct preclinical translation while working collaboratively with other institutions to apply these laboratory findings to new and better treatments.
\end{itemize}
In light of the information presented above and the fact that 7 Basic Laboratories is a research-focused organization with a sole mission to investigate and study various cancer treatment methods, we have decided to exclude them from the final list of organizations that offer treatment services to people who have cancer. This decision is based on our understanding that 7 Basic Laboratories is not equipped or staffed to provide direct medical care to cancer patients, and therefore would not be suitable for inclusion in the list of organizations offering treatment services.
In view of this information, the process of assigning medical centers to cancer patients who are in need of medical services will be limited to only 64 active centers that are affiliated with the National Cancer Institute (NCI) across the United States. These centers have been carefully selected based on their reputation, expertise, and proven track record in providing high-quality cancer treatment to patients. They are equipped with the necessary medical personnel, technology, and resources to offer a wide range of treatment options and support services to cancer patients. By limiting the list of organizations offering treatment services to these 64 active centers, we aim to ensure that cancer patients receive the best possible care and support during their journey toward recovery.\\
Patients can receive a certain level of care from each of these centers, according to their individual capacities. The term \textbf{Staffed Bed} is used in the lingo of health management and treatment to refer to this capacity. According to~\cite{AHDSB} it is defined as "the number of beds available for use by patients at the end of the cost reporting period.  A bed means an adult bed, pediatric bed, birthing room, or newborn bed maintained in a patient care area for lodging patients in acute, long-term, or domiciliary areas of the hospital. Beds in the labor room, birthing room, postanesthesia, postoperative recovery rooms, outpatient areas, emergency rooms, ancillary departments, nurses and other staff residences, and other such areas are regularly maintained and utilized for only a portion of the stay of patients (primarily for special procedures or not for inpatient lodging) are not termed a bed for these purposes." Moreover, it can be described as a hospital bed provided by the  National Health Service (NHS) that can either be used for overnight stays or for day cases. It could be a bed that has been reserved from the specialty bed complement, a bed that has been borrowed from another specialty or significant facility, or it could even be a temporary bed~\cite{SB}. Therefore, the primary objective of this project is to place patients in each of these empty beds that have been staffed. Regarding this matter, it is essential to make certain that none of them are vacant at any given time.
Moreover, there is a ranking dataset about these centers available in the “USNEWS”~\cite{USNEWS}. So, in this project, we will provide a recommendation system that will suggest the best Cancer Center for each patient based on their risk score, income, and Cancer Center availability. According to~\cite{Bristow2015} “Increased access to NCI-CCCs through regional concentration of care may be a mechanism to improve clinical outcomes”.\\

This problem can be divided into two different sub-problems, at first using Survival Analysis methods, the Risk Score ($rs_i$) for each patient will be calculated, and based on the predefined Risk Threshold, ($Tr_s$), a subset of patients will be selected for the next step which is a type of stable matching.
In order to formulate the second part of the problem as a graph-theoretic one, the following variables should be considered:
\begin{itemize}
    \item 	Patient set is defined as $P=\{p_i \mid i \in N,1<i<n\}$
	\item Each member of the patient set has a non-zero integer value for these variables \{$lp_i$, $a_i$, $rs_i$\}. Where $l_i$, $a_i$, and $rs_i$ are locations, annual income, and risk score respectively.
	\item Cancer Center set is defined as $C=\{c_j \mid j \in N,1<j<m\}$
	\item Each cancer center has a non-zero integer value for these variables \{$tc_j$, $sb_j$, $lc_j$\}. Where $tc_j$, $sb_j$, and $lc_j$ are treatment costs, the number of available staffed beds, and location.

\end{itemize}
There are three important points that should be considered in the problem formulation:
\begin{itemize}
    \item 	Patient risk score ($rs_i$) should always be equal to or greater than the predefined Risk Threshold ($Tr_s$).
	\item For each patient we define Affordable Cost which is defined as a certain portion of each patient’s income. This value can be calculated based on a percentage of the annual income. So, we will have:
\begin{equation}
\label{ac}
  ac_i= a_i*x\%  
\end{equation}
Where $x$ will be defined at the beginning of running the algorithm.

\item We assume that all Cancer Centers provide sufficient treatment for all cancer sites.
\item Alongside with Risk Score Threshold ($T_{rs}$), one other threshold should be considered:
    \begin{itemize}
        \item Accessible Distance ($T_{ad}$), which is defined as the maximum distance between each patient’s living location and the cancer center’s location.
        \item Where $T_{ad}$ will be defined at the beginning of running the algorithm.
    \end{itemize}
\end{itemize}
After calculating the values for the above-mentioned variables, a bi-partite graph can be defined with the following definition:
\begin{itemize}
    \item 	The health center recommender graph, includes pairwise non-adjacent vertex sets P and S as we defined earlier. Each $p$ is adjacent to a staffed bed in one of the cancer centers set, $c_j$, if and only if the treatment cost is affordable ($ac_i<tc_j$) and the location is accessible ($|lc_j-lp_i|<T_{ad}$).
	\item We create the preference list for each patient based on:
         \begin{itemize}
            \item 	First, make the preference list using a sorted list of all $T_{ad}$ for each patient (Bucket Sort which takes O(n) time).
	        \item Secondly, removing centers where  $ac_i>tc_j$.
        \end{itemize}
\end{itemize}
Based on this definition, the final problem is to find a stable match in this graph based on the pre-defined conditions and limitations.\\
The following section will provide an explanation of the algorithm that was developed to accomplish this goal. The preference list in this stable matching problem is actually made based on distance and cost for patients, as well as the risk score for centers. This is an essential fact that must not be overlooked (actually staffed beds). Because of this, it's possible that a patient won't be referred to a particular center because of the travel time. Patients also have the right to decline an offer being made to them. The ultimate objective is to guarantee that every available bed has an occupant. Consequently, the algorithm needs to be run repeatedly until there is no vacancy left in the bed.
\section{Algorithm}
\label{ALG}
The final algorithm for the assignment part of the project will consist of two nested loops; a) providing a preliminary list based on the pre-defined variables (inner loop) and b) running the algorithm again based on updated lists (outer loop). In the outer loop, accepted assigns are removed from the waiting list, and new patients are added to the list based on the number of staffed beds that are still available. It means the outer loop is a simple loop that just updates the list of patients (P set) and Cancer Centers (C) and runs the inner loop.
Overall, the main algorithm consists of these steps:
\begin{itemize}
    \item Calculating the risk score for each patient and creating a non-increasing ranking list based on these scores and storing it in a waiting list.
    \item Creating the list of Cancer Centers (Locations, No. of staffed beds, and Costs).
    \item Selecting the first sb (number of total available staffed beds) patients from the waiting list.
    \item Generating the preference list for each patient based on their location and income.
    \item Running the Stable Matching algorithm and offering suggestions to patients.
    \item Refreshing the patients and centers list by removing the assigned ones.
    \item Re-running the algorithm.
\end{itemize}
The inner loop is described as what follows:

\subsection{Input:}
\begin{enumerate}
    \item 	A list $C$ (Cancer Centers), with each $c \in C$ associated with three integers $tc_j$, $sb_j$ and $lc_j$ (treatment cost, number of available staffed bed and location). 
	\item A list $P$ (patients), with each $p \in P$ associated with three numbers $l_i$, $a_i$, and $rs_i$ (location, annual income, and risk score respectively), and with a sorted list of a subset $c_{p}$ of $C$, where the notation $c \succ_p c'$ is used if $p$ ranks $c$ above $c'$.
\end{enumerate}

\subsection{Data structures:}
\begin{enumerate}
    \item  $OriginalList$: A list of all cancer patients sorted based on the risk score
    \item  $WaitList$: A list of patients, sorted by their risk score from the highest to the lowest (those whose processing has not yet started). 
    \item  $ProcessList$: A list of patients (those who are currently processed). 
    \item  $UnassignedList$: A list of patients (those who will not get a bed) 
    \item  $CenterStatus$: An array of size $|C|$ that contains for each $c \in C$ : 
        \begin{enumerate}
            \item $PatientList[c]$: A list of patients (those currently matched to $c$), sorted by risk score from the highest to the lowest 
            \item $StaffedBed[c]$: $\{ sb - |PatientList[c]| \}$ (the current number of empty staffed beds in $c$).
        \end{enumerate} 
    \item $PatientStatus$: An array of size $|S|$ that contains for each $p \in P$:
        \begin{enumerate}
            \item $Status[p]$: A symbol (the current status of p ) which belongs to one of the following: “$W$” ($p$ is in $WaitList$), “$P0$” ($p$ is in $ProcessList$ but is not assigned), “$P1$” ($p$ is in $ProcessList$ and is assigned) or “$O$” ($p$ is in $UnassignedList$). 
            \item $CenterList[p]$: A sorted sublist of $c_p$ (Cancer Centers which are still open for $p$).
        \end{enumerate}
\end{enumerate}

\subsection{Initialization phase:}
\begin{enumerate}
    \item Move all patients from $UnassignedList$ to $WaitList$
    \item Extract the first $|sb - |WaitList||$ number of patients form the $OriginalList$ and add them to $WaitList$
	\item For each patient $p \in P$, set [$Status[p]$ $\leftarrow$ $W$] and [$CenterList[p]$ $\leftarrow$ $c_p$]
	\item For each $c \in C$ , set [$PatientList[c]$ $\leftarrow$ $0$] and [$StaffedBed[c]$ $\leftarrow$ $sb$ ]. 
	\item Set [$UnassignedList$ $\leftarrow$ $0$]. 
	\item Set $x$.
	\item Set $T_{ad}$.
	\item Move all patients from $P$ into $WaitList$ and sort them by their risk score. 
	\item Move the top $min\{ |S|, sb \}$ patients from $WaitList$ to $ProcessList$ and change their status from $W$ to $P0$.
\end{enumerate}

\subsection{Iteration phase:}
\begin{enumerate}
    \item \textit{\textbf{If}} there is a patient in $ProcessList$ with status $P0$: Let $p$ be any such patient.\\\hspace*{10mm}Go to Step $2$.\\
          \textit{\textbf{Else}} Go to the Termination Phase.\\\\
    \item \textit{\textbf{If}} $CenterList[p]$ is not empty: Let $c$ be the top element in $CenterList[p]$.\\\hspace*{10mm}Go to Step $3$.\\
          \textit{\textbf{Else}} $CenterList[p]$ is empty: Move $p$ from $ProcessList$ into $UnassignedList$ and change $Status[p]$ from $P0$ to $O$.\\\\
          \textit{\textbf{If}} $WaitList$ is not empty: Move the patient from the top of $WaitList$, say patient $p"$, to $ProcessList$ and change $Status[p"]$ from $W$ to $P0$.\\\hspace*{10mm}Go to Step $1$.\\
          \textit{\textbf{Else}} $WaitList$ is empty:\\\hspace*{10mm}Go to Step $1$.\\\\
    \item \textit{\textbf{If}} $StaffedBed[c]$ $>$ $0$: Add patient $p$ to the right position in $PatientList[c]$ (according to his/her risk score), change $Status[p]$ from $P0$ to $P1$ and reduce $StaffedBed[c]$ by $1$.\\\hspace*{10mm}Go to $1$.\\
          \textit{\textbf{Else If}} $StaffedBed[c]$ $=$ $0$ and the risk score of p is less than that of the last patient in $PatientList[c]$: Remove c from $CenterList[p]$.\\\hspace*{10mm}Go to step $2$.\\
          \textit{\textbf{Else}} $StaffedBed[c]$ $=$ $0$ and the risk score of p is higher than that of the last patient in $PatientList[c]$ say $p'$: Remove $p'$ from $PatientList[c]$ and change $Status[p']$ from $P1$ to $P0$. Add patient $p$ to the right position in $PatientList[c]$ (according to the risk score) and change $Status[p]$ from $P0$ to $P1$.\\\hspace*{10mm}Go to step $1$.\\ 

\end{enumerate}

\subsection{Output Phase}
\begin{itemize}
    \item The output is the triplet (M, W, U) where M = {(p, c) : p appears in PatientList[c]}, W consists of the patients in WaitList (those with Status[p] = W) and U consists of the patients in UnassignedList (those with Status[p] = U).
    \item M will be offered to patients:
    \begin{itemize}
        \item Patients who approved suggestions will be removed from OriginalList
        \item Patients who did not approve suggestions will be added to UnassignedList
    \end{itemize}
\end{itemize}

\section{Experimental Results and Time Complexity}
\label{Analysis}
\subsection{Experimental Results}
To implement the experimental phase, we have used the patient information available in the SEER dataset~\cite{SEER}. Also, medical centers have been collected exactly according to the list announced by NCI.
The Surveillance, Epidemiology, and End Results (SEER) Program~\cite{SEER} provides information on cancer statistics in an effort to reduce the cancer burden among the U.S. population. SEER is supported by the Surveillance Research Program (SRP) in NCI's Division of Cancer Control and Population Sciences (DCCPS). Using the available datasets in this database, we can calculate Survival Function and Hazard Function for each patient. In order to implement the algorithm, we extracted the information of patients with Breast Cancer in different states. Table~\ref{table:SEER} shows the details about the patients in different states.\\Because the SEER dataset does not include information about an individual's income, we used a number that was derived from the average income for each state that is provided in the SEER population dataset. This is an important point to keep in mind.

\renewcommand{\thetable}{\arabic{table}}
\begin{table*}[ht]
\centering
\newcommand{\myline}{\cline{1-6}}
\begin{tabular}{@{}||l ||l| l| l| l |l ||@{}}
\myline
\textbf{Dataset Name} & \textbf{Dataset Dim.} & \textbf{No. Events} & \textbf{No. Censoring} & \textbf{Pre-Coding Size (KB)} &  \textbf{Post-Coding Size (KB)} \\
 \myline
 \myline
Overall & 1,008,976&	125,309	(12\%)&	883,667	(88\%)&	360,103	&84,618\\
California & 409,880&	49,839	(12\%)	&360,041	(88\%)	&146,553&	34,140\\
Connecticut & 51,372	&	5,223	(10\%)	&	46,149	(90\%)	&	18,537	&	4,226\\
Georgia & 107,623	&	14,570	(14\%)	&	93,053	(86\%)	&	37,379	&	8,863\\
Hawaii & 17,515	&	1,616	(9\%)	&	15,899	(91\%)	&	6,415	&	1,500\\
Iowa & 38,215	&	4,390	(11\%)	&	33,825	(89\%)	&	13,731	&	3,149\\
Kentucky & 53,522	&	7,334	(14\%)	&	46,188	(86\%)	&	18,716	&	4,406\\
Louisiana & 53,467	&	8,110	(15\%)	&	45,357	(85\%)	&	18,729	&	4,403\\
New jersey & 119,271	&	15,124	(13\%)	&	104,147	(87\%)	&	42,618	&	9,942\\
New Mexico & 22,014	&	2,902	(13\%)	&	19,112	(87\%)	&	7,825	&	1,801\\
Utah & 21,854	&	2,729	(12\%)	&	19,125	(88\%)	&	7,607	&	1,798\\
\myline
\end{tabular}
\caption{Descriptive Statistics and Information about SEER Breast Cancer. The total number of all states is less than the General because some rows, in general, belong to specific cities that have not appeared in the study.}
\label{table:SEER}
\end{table*}

Regarding the details about the NCI-Designated Cancer Centers, the list has been derived from~\cite{NCI}. However, the number of staffed beds is not available in~\cite{NCI}, so we used~\cite{covMap} and~\cite{CalFac} which contain the number of staff beds in the US healthcare facilities and merge the numbers with the first list. The final list containing all needed information about NCI-Designated Cancer Centers is shown in table~\ref{table:centers}.

\begin{table*}[t!]\centering
\begin{tabular}{|l|l|l|l|l|l|}
\hline
\textbf{No}. & \textbf{Center Name}                                                                        & \textbf{City}            & \textbf{State}                & \textbf{Type} & \textbf{SB$^*$} \\ \hline \hline
1   & O'Neal Comprehensive Cancer Center                                                  & Birmingham      & Alabama              & 3C$^{*}$   & 1063         \\ \hline
2   & Arizona Cancer Center                                                               & Tucson          & Arizona              & 3C   & 479          \\ \hline
3   & Chao Family   Comprehensive Cancer Center                                           & Orange          & California           & 3C   & 352          \\ \hline
4   & Stanford Cancer   Institute (SCI)                                                   & Stanford        & California           & 3C   & 447          \\ \hline
5   & City of Hope   Comprehensive Cancer Center                                          & Duarte          & California           & 3C   & 217          \\ \hline
6   & UC Davis   Comprehensive Cancer Center                                              & Sacramento      & California           & 3C   & 598          \\ \hline
7   & Jonsson Comprehensive   Cancer Center                                               & Los Angeles     & California           & 3C   & 214          \\ \hline
8   & Moores Comprehensive   Cancer Center                                                & La Jolla        & California           & 3C   & 711          \\ \hline
9   & UCSF Helen Diller   Family Comprehensive Cancer Center                              & San Francisco   & California           & 3C   & 170          \\ \hline
10  & USC Norris   Comprehensive Cancer Center                                            & Los Angeles     & California           & 3C   & 60           \\ \hline
11  & University of   Colorado Cancer Center                                              & Aurora          & Colorado             & 3C   & 650          \\ \hline
12  & Yale Cancer Center                                                                  & New Haven       & Connecticut          & 3C   & 1279         \\ \hline
13  & Georgetown Lombardi   Comprehensive Cancer Center                                   & Washington      & District of Columbia & 3C   & 393          \\ \hline
14  & Moffitt Cancer Center                                                               & Tampa           & Florida              & 3C   & 295          \\ \hline
15  & Sylvester   Comprehensive Cancer Center                                             & Miami           & Florida              & 2C$^{*}$   & 40           \\ \hline
16  & Winship Cancer   Institute                                                          & Atlanta         & Georgia              & 3C   & 475          \\ \hline
17  & University of Hawaii   Cancer Center                                                & Honolulu        & Hawaii               & 2C   & 33           \\ \hline
18  & Robert H. Lurie   Comprehensive Cancer Center                                       & Chicago         & Illinois             & 3C   & 312          \\ \hline
19  & The University of   Chicago Comprehensive Cancer Center                             & Chicago         & Illinois             & 3C   & 682          \\ \hline
20  & Indiana University   Melvin and Bren Simon Comprehensive Cancer Center              & Indianapolis    & Indiana              & 3C   & 1226         \\ \hline
21  & Holden Comprehensive   Cancer Center                                                & Iowa City       & Iowa                 & 3C   & 716          \\ \hline
22  & The University of   Kansas Cancer Center                                            & Kansas City     & Kansas               & 2C   & 848          \\ \hline
23  & Markey Cancer Center                                                                & Lexington       & Kentucky             & 2C   & 901          \\ \hline
24  & Sidney Kimmel   Comprehensive Cancer Center                                         & Baltimore       & Maryland             & 3C   & 411          \\ \hline
25  & University of   Maryland Marlene and Stewart Greenebaum Comprehensive Cancer Center & Baltimore       & Maryland             & 3C   & 829          \\ \hline
26  & DanaFarber/Harvard   Cancer Center                                                  & Boston          & Massachusetts        & 3C   & 30           \\ \hline
27  & The Barbara Ann   Karmanos Cancer Institute                                         & Detroit         & Michigan             & 3C   & 123          \\ \hline
28  & University of   Michigan Rogel Cancer Center                                        & Ann Arbor       & Michigan             & 3C   & 931          \\ \hline
29  & Masonic Cancer Center                                                               & Minneapolis     & Minnesota            & 3C   & 778          \\ \hline
30  & Mayo Clinic Cancer   Center                                                         & Rochester       & Minnesota            & 3C   & 1115         \\ \hline
31  & Alvin J. Siteman   Cancer Center                                                    & St. Louis       & Missouri             & 3C   & 430          \\ \hline
32  & Fred and Pamela   Buffett Cancer Center                                             & Omaha           & Nebraska             & 2C   & 345          \\ \hline
33  & Darthmouth Cancer   Center                                                          & Lebanon         & New Hampshire        & 3C   & 211          \\ \hline
34  & Rutgers Cancer   Institute of New Jersey                                            & New Brunswick   & New Jersey           & 3C   & 150          \\ \hline
35  & University of New   Mexico Cancer Research and Treatment Center                     & Albuquerque     & New Mexico           & 3C   & 410          \\ \hline
36  & Montefiore Einstein   Cancer Center                                                 & Bronx           & New York             & 2C   & 380          \\ \hline
37  & Memorial   SloanKettering Cancer Center                                             & New York        & New York             & 3C   & 498          \\ \hline
38  & Roswell Park   Comprehensive Cancer Center                                          & Buffalo         & New York             & 3C   & 133          \\ \hline
39  & Herbert Irving   Comprehensive Cancer Center                                        & New York        & New York             & 3C   & 236          \\ \hline
40  & Tisch Cancer   Institute                                                            & New York        & New York             & 2C   & 411          \\ \hline
41  & Laura and Isaac   Perlmutter Cancer Center at NYU Langone Health                    & New York        & New York             & 3C   & 110          \\ \hline
42  & Duke Cancer Institute                                                               & Durham          & North Carolina       & 3C   & 76           \\ \hline
43  & Wake Forest Baptist   Comprehensive Cancer Center                                   & WinstonSalem    & North Carolina       & 3C   & 130          \\ \hline
44  & UNC Lineberger   Comprehensive Cancer Center                                        & Chapel Hill     & North Carolina       & 3C   & 105          \\ \hline
45  & Case Comprehensive   Cancer Center                                                  & Cleveland       & Ohio                 & 3C   & 63           \\ \hline
46  & The Ohio State   University Comprehensive Cancer Center                             & Columbus        & Ohio                 & 3C   & 344          \\ \hline
47  & Stephenson Cancer   Center                                                          & Oklahoma City   & Oklahoma             & 2C   & 72           \\ \hline
48  & Knight Cancer   Institute                                                           & Portland        & Oregon               & 3C   & 111          \\ \hline
49  & Abramson Cancer   Center                                                            & Philadelphia    & Pennsylvania         & 3C   & 680          \\ \hline
50  & UPMC Hillman Cancer   Center                                                        & Pittsburgh      & Pennsylvania         & 3C   & 93           \\ \hline
51  & Fox Chase Cancer   Center                                                           & Philadelphia    & Pennsylvania         & 3C   & 100          \\ \hline
52  & Sidney Kimmel Cancer   Center at Thomas Jefferson University                        & Philadelphia    & Pennsylvania         & 2C   & 124          \\ \hline
53  & Hollings Cancer   Center                                                            & Charleston      & South Carolina       & 2C   & 239          \\ \hline
54  & St. Jude Children's   Research Hospital                                             & Memphis         & Tennessee            & 3C   & 212          \\ \hline
55  & VanderbiltIngram   Cancer Center                                                    & Nashville       & Tennessee            & 3C   & 281          \\ \hline
56  & Dan L Duncan   Comprehensive Cancer Center                                          & Houston         & Texas                & 3C   & 63           \\ \hline
57  & Mays Cancer Center at   UT Health San Antonio                                       & San Antonio     & Texas                & 2C   & 78           \\ \hline
58  & Harold C. Simmons   Comprehensive Cancer Center                                     & Dallas          & Texas                & 3C   & 243          \\ \hline
59  & The University of   Texas MD Anderson Cancer Center                                 & Houston         & Texas                & 3C   & 674          \\ \hline
60  & Huntsman Cancer   Institute                                                         & Salt Lake City  & Utah                 & 3C   & 412          \\ \hline
61  & Massey Cancer Center                                                                & Richmond        & Virginia             & 2C   & 119          \\ \hline
62  & University of   Virginia Cancer Center                                              & Charlottesville & Virginia             & 3C   & 701          \\ \hline
63  & Fred   Hutchinson/University of Washington Cancer Consortium                        & Seattle         & Washington           & 3C   & 282          \\ \hline
64  & University of   Wisconsin Carbone Cancer Center                                     & Madison         & Wisconsin            & 3C   & 493          \\ \hline
\end{tabular}
\caption{All NCI-Designated Cancer Centers with their city name, state, type, and the number of staffed beds are shown in the table. \\ * \textbf{3C}: Comprehensive Cancer Centers, \textbf{2C}: Cancer Centers, and \textbf{SB}: Staffed Beds.}
\label{table:centers}
\end{table*}
 It is worth noting that since centers of type "Basic Laboratory" are not providing any treatment, we removed them from the list. Furthermore, the total number of staffed beds in all 64 centers (71 - 6 basic laboratories) is $25,387$. In order to implement the algorithm, we assumed that $20\%$ of the staffed beds are available for the assignment process which is $5,077$. 
Same as the SEER dataset, the exact treatment cost is not available for each cancer center. So we calculated an average treatment cost for each state based on~\cite{Blumen2016},~\cite{F1},~\cite{F2}, and~\cite{F3}.

In the problem formulation, the distance is calculated based on the patient's state based on these:
\begin{itemize}
    \item Distance for residency state is 0.5.
    \item Distance from neighbor states is 1.
    \item Distance increases by 1 for each neighbor of neighbor and this goes on.
\end{itemize}

The algorithm has been run based on the above-mentioned variables and we assumed that offers have been accepted by the patients. To make the calculations for Eq.~\ref{ac}, we define x (the percentage of annual income that patients can pay for treatment) as $25\%$. In addition to that, the cutoff point for the distance is currently set to 3.
In order to successfully assign all of the available staffed beds to patients, the outer loop has been executed a total of 4 times.
It should come as no surprise that altering these variables and thresholds will result in a shift in the total number of times the algorithm will be executed. In subsequent works, we will make an effort to select appropriate values for these variables by basing our decisions on findings from additional research conducted on this topic.

\subsection{Time Complexity}
We were unable to make a fair comparison between this algorithm and other algorithms because it was the first of its kind.
Concerning the difficulty of the task in terms of the amount of time it will take, we are aware that the phase that will take the most time will be the assignment (stable matching), which will never exceed the complexity level of $O(nm)$, where $n$ is the number of patients and $m$ is the number of staffed beds in cancer centers. Moreover, among all the sorting processes that have been done before running the main stable matching algorithm, risk score sorting has taken the most time and space. However, since scores are between $0$ and $1.0$ and the increase rate is $0.01$, they have been sorted using Bucket Sort where the number of buckets ($100$) is much lower than the number of patients (at least $1,000,000$). It is worth noting that since the risk score is rarely less than $0.50$, so the number of buckets will be around $50$ in most cases.

It is also important to note that all of the codes and packages will be published on GitHub in order to make them accessible to others so that they can study them and work on improving them.

\section{Conclusion}
\label{CL}
The treatment of cancer is one of the most discussed issues in the realm of contemporary public health research. One of the primary concerns of both the general public and the government is the development of the most effective cancer treatment at the most affordable price. This is due to the fact that the number of people diagnosed with cancer increases on an annual basis. There are sixty-four cancer centers in the United States that have received the NCI Designation. In light of advancements in patient management systems and smart city technologies \cite{seidi2019pv, seidi2017vid}, our project takes a novel approach by recasting this problem as a stable matching problem. We developed an algorithm for determining which cancer treatment center is the most appropriate for each individual patient, leveraging insights from these advanced systems to enhance patient-centric care and optimize resource allocation in healthcare.

\section{Future Works}
\label{FW}
In this work, we only considered one type of treatment in each center, but because there are multiple treatments that can be provided in the real world based on the center and the patient's needs, future work could map this problem to a 3-dimensional stable matching.\\
In addition, a more effective and comprehensive version of this project might include compiling a master list based on the order in which the cancer treatment centers are listed.\\
Moreover, altering the values for the thresholds, like affordable cost, distance, and risk score, will result in a shift in the total number of times the algorithm will be executed. In future works, we will try to choose the right values for these variables by basing our choices on the results of more research on this topic.\\
In end, the problem environment can be extended to a version where patients and beds come and go based on a specific distribution, and the algorithm should be run according to that. The outer loop of the final algorithm should be chosen to run based on this distribution.

\bibliographystyle{IEEEtran}
\bibliography{ref}


\end{document}